\RequirePackage{fix-cm}
\documentclass[smallextended]{svjour3}       % onecolumn (second format)
\smartqed  % flush right qed marks, e.g. at end of proof
\usepackage{graphicx}
\def\makeheadbox{{%
\hbox to0pt{\vbox{\baselineskip=10dd\hrule\hbox
to\hsize{\vrule\kern3pt\vbox{\kern3pt
\hbox{\bfseries [Arxiv]}
\hbox{This is a pre-peer-review version of this article.}
\kern3pt}\hfil\kern3pt\vrule}\hrule}%
\hss}}}

\usepackage{lineno,hyperref}
\usepackage{bm}% bold math
\usepackage{amsmath}
\usepackage{amssymb}
\usepackage{bbold}
\usepackage{cite}
\def\v#1{{\bf#1}}
\def\be{\begin{equation}}
\def\ee{\end{equation}}
\def\bea{\begin{eqnarray}}
\def\eea{\end{eqnarray}}

\def\<{\langle}
\def\>{\rangle}

\begin{document}

\title{Exactly solvable SIR models, their extensions and their application to sensitive pandemic forecasting}

\titlerunning{Exactly solvable SIR models}        % if too long for running head

\author{E. Sadurn\'i        \and
        G. Luna-Acosta %etc.
}


\institute{E. Sadurn\'i \at
              Instituto de F\'isica,Benem\'erita Universidad Aut\'onoma de Puebla,
Apartado Postal J-48, 72570 Puebla, M\'exico \\
              \email{emerson.sadurni@gmail.com}           %  \\
           \and
           G. Luna-Acosta \at
              Instituto de F\'isica,Benem\'erita Universidad Aut\'onoma de Puebla,
Apartado Postal J-48, 72570 Puebla, M\'exico \\
             \email{glunaacosta@gmail.com}
}

\date{Received: date / Accepted: date}

\maketitle

\begin{abstract}
The classic SIR model of epidemic dynamics is solved completely by quadratures, including a time integral transform expanded in a series of incomplete gamma functions. The model is also generalized to arbitrary time-dependent infection rates and solved explicitly when the control parameter depends on the accumulated infections at time $t$. Numerical results are presented by way of comparison. Autonomous and non-autonomous generalizations of SIR for interacting regions are also considered, including non-separability for two or more interacting regions. A reduction of simple SIR models to one variable leads us to a generalized logistic model, Richards model, which we use to fit Mexico's  COVID-19 data up to day number 134. Forecasting scenarios resulting from various fittings are discussed.  A critique to the applicability of these models to current pandemic outbreaks in terms of robustness is provided. Finally, we obtain the bifurcation diagram for a discretized version of Richards model, displaying period doubling bifurcation to chaos. 
\end{abstract}

\section{Introduction}

\paragraph{History} Epidemiological models based on first order, linear and multidimensional differential equations can be traced back to Daniel Bernoulli \cite{dietz2002}. The famous SIR (Susceptible Infected Removed) and SIRD (plus Dead) systems were developed by Sir Ronald Ross nearly a hundred years ago; see \cite{weiss2013}. The former entails the use of non-linear terms in the model expressions \cite{hethcote1989, hethcote2000, hethcote2009} with the aim of describing interactions between S, I and R populations, together with the introduction of obvious critical points at vanishing populations of infected or susceptible individuals.
Since the introduction of those models in the past century, the possibility of establishing control mechanisms based on predictions has been part of public health policies. The relevance of sufficiently simple contagion dynamics of this kind cannot be underestimated, particularly in recent times. There has been a continued interest in this subject as shown by \cite{anderson1991, hu2012, eckalbar2011}, but modern and ongoing research based on SIR models can be found in \cite{kumar2020, roy2020, sjodin2020}.

\paragraph{Motivation} It may appear at first sight that the required equilibrium points for any sensible dynamical model of epidemics would impose non-linearities which are insurmountable for an exact analysis. It is well known that some qualitative features of the SIR models can be extracted from the defining equations themselves, but the precise evolution of populations in terms of known functions necessitates explicit solutions. It is also desirable that the resulting expressions be amenable to numerical evaluations to avoid the use of robust numerical algorithms for multivariable first order problems. For our purposes, it is important to know whether non-linearities hamper full integrability. Integrability in the sense \cite{encyclopedia} that there exist a sufficient number of  integration constants that allow  separability of  multivariable differential equations, which is more general than the existence of explicit solutions. The lack of this quality would relegate the study to numerically obtained flows and unpredictability when the system is in a chaotic regime. Fortunately, in the context of SIR models we will be able to show that the standard versions of SIR models allow separability and a reduction to quadratures (in mechanical systems, these would be the integrals for periods or lapses). Moreover, we shall see that the associated integral transforms allow evaluation under reasonable assumptions and approximations. The integrability of these models is another motivating factor to carry out  future work in the direction of determining and designing appropriate policies to ameliorate the effects of an epidemic.

\paragraph{Goals and results} Separable systems fall into the category of integrable models. In this paper we shall address the conditions for separability in in the aforementioned SIR model. It is sometimes mentioned \cite{weiss2013} that either, explicit solutions of SIR do not exist or they are unknown. We assert that this is not entirely correct: we shall see from the full integrability of the model that a last quadrature related to the lapse function or time interval can be found. The latter is given in terms of known functions, and it solves the problem explicitly in its entirety by reverse substitution into all previous expressions for population functions S, I and R. Incomplete gamma functions will appear in the results. The only caveat here is that such lapse or period integral cannot be put in terms of {\it elementary\ }functions and only an infinite series of them can be offered. However, their form is sufficiently amenable for numerical evaluation.

In the realm of non-autonomous systems and the application of policies in real time, we shall address a generalization of SIR with a sliding infection parameter --e.g. controllable by quarantine-- and show that this problem is integrable as well if the total population is conserved and whenever the control parameter of infection depends explicitly on accumulated cases of infection at a given time. A generalization of SIR to interacting populations will be given in this work and we will see that the number of separation constants does not match the number of degrees of freedom and then, only in this case, we shall have a lack of explicit solutions.

As an important application of our results, we shall address the problem of curve fitting in the context of COVID-19 acquired data in the case of Mexico, during the year 2020. A simple generalization of the logistic model, the Richards model,  may capture the general behaviour of the accumulated infection curve, but errors in data acquisition may give rise to poor forecasting a few weeks ahead; the stability of the system in terms of data uncertainties will be studied numerically and analytically.

\paragraph{Structure of the paper} In section \ref{sec:1} we present a self-contained formulation of SIR and SIRD models, together with their variations in both solvable and non-separable cases. In subsection \ref{sec:1.1} explicit solutions are presented, including a lapse integral in terms of gamma functions and equilibrium conditions. Subsection \ref{sec:1.2} contains a treatment with variable infection rates. The case of two interacting regions is discussed in \ref{sec:1.3} and in subsection \ref{sec:1.4} the general case of many interacting regions and total distributions is presented. Section \ref{sec:2} is devoted to an application of Richards model to the case of Mexico's Covid19 outbreak in the year 2020. In Subsection \ref{sec:2.1} the explicit solutions of the model are presented, together with an analysis of robustness under data fluctuations. Subsection \ref{sec:2.2} discusses the various forecasting scenarios resulting from curve fittings to COVID-19 in Mexico. In Subsection\ref{sec:2.3} we study the dynamics of a discretized version of Richards model and in Section \ref{sec:3} we draw our conclusions. 

\section{SIR and SIRD models \label{sec:1}}

The total population $P(t)$ is a differentiable function that can be divided into susceptible individuals $S(t)$, infected individuals $I(t)$ and removed individuals $R(t)$. It is assumed that $I$ can infect others contained in $S$, but $R$ is inactive. This removed population may contain cured and dead cases as $R=C+D$, where $C$ and $D$ are also considered inactive. It is also assumed that neither sources nor sinks are present in $P$, so this quantity is conserved; some generalizations relax this supposition. In this manner, $C$ and $D$ are not part of the dynamical system and they can always be recovered by a statistical treatment of death percentage $D=qR$, $C=pR$ such that $p+q=1$. This leaves us with SIR instead of SIRD. The velocity at which contagion occurs can be estimated in controlled (experimental) situations when a few members of the population come into contact. For two individuals sharing a common space of a given radius the initial values $R=0$, $S=1$, $I=1$, $P=2$ start a process for which at time $\Delta t_i$ we end up with $S=0$ and $I=2$. Because of proximity, the quantity $\Delta t_i$ is a function of population density. Since the new infected population is 2, $\Delta S = -1$ and we have $\Delta S / \Delta t_i = -1/\Delta t_i \equiv -\delta$, so our new infection constant is reciprocal to the infection time. This situation can be scaled to more infected and more susceptible individuals; if a number $S_0$ shares the space with $I_0 = 1$, they will pass to the next category in the same time $\Delta t_i$ and if more infected individuals are active $I_0 > 1$ the probability of contagion increases proportionally. This establishes, in the limit of small time steps, the law

\bea
\frac{dS}{dt} = - \delta SI.
\label{1}
\eea   
The constant $\delta$ so defined can be corrected in experiments involving larger populations. Similarly, the infected $I$ suffer a change $\Delta I$ which comes from two contributions: a depletion term due to removal (death or recovery) modulated by  some coefficient $\alpha$  in a time $\Delta t_r$ which is proportional to $I$ and a gain in a time $\Delta t_i$ controlled by a coefficient $\beta$ which competes with the loss in $S$. We have 

\bea
\frac{dI}{dt} = \beta SI - \alpha I.
\label{2}
\eea
Finally, the removed population can be fed only from $I$, assuming there is no preventive cure to turn $S$ into $R$ (specifically $C$) directly. Therefore, the growth velocity of $R$ is proportional to $I$ with a constant $\gamma$ also measured by the reciprocal time of the removal process:

\bea
\frac{dR}{dt} = \gamma I.
\label{3}
\eea
Even in the case where the proportionality factors were unkown, the conservation of $P=S+I+R$ forces $\delta=\beta,\gamma=\alpha$, as can be checked by adding the expressions (\ref{1}), (\ref{2}) and (\ref{3}) with $dP/dt=0$.  The analysis can be simplified  by  redefining  the  time scale,  $\tau = \beta t$, leading to only one effective control parameter $\kappa$ in the model, i.e. 

\bea
\frac{dI}{d\tau} = (S-\kappa)I, \quad \frac{dS}{d\tau} = -SI, \quad \frac{dR}{d\tau} = \kappa I,
\label{4}
\eea
where 

\bea
\kappa \equiv \frac{\Delta t_i}{\Delta t_r}=\frac{\alpha}{\beta}
\label{5}
\eea
is the infection-to-recovery time ratio. The system (\ref{4}) is autonomous and it is exactly solvable, as we now show.

\subsection{One Spatial Region with a piecewise definition of infection rates \label{sec:1.1}}

Let us consider $\kappa$ constant in a fixed time interval. The interval could be taken as $[0,\infty)$ for asymptotic analysis. The relations (\ref{4}) imply the quotients

\bea
\frac{dI}{dS}= \frac{\kappa -S}{S}, \quad \frac{dS}{dR}=- \frac{S}{\kappa},
\label{6}
\eea
which constitute separable ordinary differential equations with solutions

\bea
S&=& S_0 e^{(R_0 - R)/\kappa}, \quad I = I_0 + S_0 + \kappa \log (S/S_0) - S, \nonumber \\  R&=& S_0 + I_0 + R_0 - I - S = R_0 - \kappa \log (S/S_0).
\label{7}
\eea
To extract the time behaviour $S(\tau), I(\tau), R(\tau)$ we eliminate $I(R)$ in the third relation of (\ref{4}) using the second relation in (\ref{7}):

\bea
I &=& I_0 + (R_0 - R) + S_0 - S_0 e^{(R_0-R)/\kappa} = P-R-S_0 e^{(R_0-R)/\kappa} \nonumber \\
\frac{dR}{d\tau} &=& \kappa \left(P-R-S_0e^{(R_0-R)/\kappa}\right).
\label{8}
\eea
The last relation above is an ordinary differential equation for $R$ alone and can be separated, thus,  it can be solved by quadrature to yield the so-called lapse integral:

\bea
\beta \Delta t = \Delta \tau = \int_{R_0}^{R} \frac{dX}{\kappa \left(P-X-S_0e^{(R_0-X)/\kappa}\right) }.
\label{9}
\eea
\paragraph{Integral transform for the lapse function} From the growth relation $dR/d\tau = \kappa I$ we know that if $\kappa \neq 0$ and $I \neq 0$ i.e., away from the trivial equilibrium point, it holds that $dR/d\tau \neq 0$ for all times in $[0,\infty)$. As a consequence (\ref{9}) defines a monotonic invertible function and $R(\tau)$ can be recovered. Furthermore, the integral transform in (\ref{9}) can be calculated in two regimes: $\kappa<1$ and $\kappa>1$. If $\kappa < 1$, then $e^{(R_0-X)/\kappa} < 1$ as here $X>R_0$. A geometric expansion of the integrand is viable and obtains  

\bea
\Delta \tau &=& \sum_{n=0}^{\infty} \frac{(-1)^{n+1}}{\kappa} \left( \frac{n S_0 e^{(R_0-R)/\kappa}}{\kappa} \right)^n 
\nonumber \\ & \times & \left[ \Gamma\left(-n, \frac{(R_0 - P)n}{\kappa} \right) - \Gamma\left(-n, \frac{(R - P)n}{\kappa} \right) \right]
\label{10}
\eea
where $\Gamma$ is the incomplete gamma function. At lowest order when $\kappa \rightarrow 0$ one gets from either Eq.(9) or Eq.(10)
\bea
R = R_0 + (I_0 + S_0) \left( 1 - e^{-\alpha \Delta t} \right),
\label{11}
\eea
 where we use $\kappa \beta= \alpha$, c.f. Eq. (\ref{5}). This states that as $\Delta t \rightarrow \infty$, $R \rightarrow P \equiv R_{\rm max}$ and all the population $P$ is transferred to the removed state, as expected from a large infection rate $\beta$. On the other hand, if  $\kappa > 1$, then $\zeta \equiv |e^{(R_0-X)/\kappa}-1|<1$ by expanding the exponential. In this case, a geometric series in $\zeta$ is plausible, and (\ref{9}) leads once more to exponential integrals conforming a series of gamma functions:

\bea
\Delta \tau &=& \frac{1}{\kappa} \sum_{n=0}^{\infty} \sum_{k=0}^{n} (-1)^{k+n} \left( \begin{array}{c} n \\ k \end{array} \right) \left( \frac{k S_0}{\kappa} \right)^n e^{k I_0/\kappa} \nonumber \\ & \times & \left[ \Gamma\left(-n, k I_0/\kappa \right) - \Gamma\left(-n, k (P-R-S_0)/\kappa \right) \right].
\label{12}
\eea
To lowest order  ($e^{(R_0-X)/\kappa}\sim 1)$, both, Eq.(8)  and  (9)  yield

\bea
R=R_0 + I_0 (1-e^{-\alpha \Delta t})
\label{13}
\eea
which implies $R \rightarrow R_0 + I_0 $ at infinity, a less drastic behavior than than the previous case (\ref{11}).  To further distinguish between these cases, we note that $\kappa$ also represents the susceptible population needed to reach the maximum infection, as can be inferred from the vanishing derivative in (\ref{4}). This means that $\kappa \gg 1$ in realistic models, where infection peaks take place and $S \gg 1$ simultaneously.

\paragraph{Equilibria} Before treating the case of jumping values of $\kappa$ during a controlled outbreak, it is important to recall the critical behaviour of the involved quantities. From (\ref{4}) we see that the infection peak occurs when $S=\kappa$, and if we use (\ref{7}) the peak value $I_{\rm max}$ results in 

\bea
I_{\rm max} = I_0 + S_0 + \kappa \left( \log (\kappa/S_0) -1 \right)
\label{14}
\eea
which is meaningful when $I_0 + S_0 \geq \kappa \left( \log (S_0/\kappa) +1 \right)$ (this is always the case if $P$ is large). If $\kappa=S_0$ then $I_{\rm max}({\rm min})= I_0 + S_0 -S_0 =I_0$ and the peak occurs trivially at the initial condition. On the other hand, as $\kappa$ increases, the function $I_{\rm max}(\kappa)$ decreases when $\kappa < S_0$; indeed, if the susceptible number $S$ decreases with the dynamics, $S_0$ is larger than the critical population $S=\kappa$ and from (\ref{14}) we find that the last two terms are negative, reducing the value of  $I_{\rm max}(\kappa)$ and explaining the famous curve flattening.

In order to recover the total deaths we look now at $R \rightarrow R_{\infty}$ as $\Delta \tau \rightarrow \infty$. By direct time integration (see Eq.(\ref{4})) , we have

\bea
R_{\infty} = R_0 + \kappa \int_{0}^{\infty} d\tau I(\tau).
\label{15}
\eea 
 
Eq. (\ref{14}) shows that $I(\tau)$ depends on $\kappa$; hence  $R_\infty$ is  also $\kappa$-dependent. From the definition of $\kappa=$ Eq. (\ref{5}) and the meanings of $\alpha$ and $\beta$, modification of either $\beta$ (by changing quarantine duration) or $\alpha$ (vaccine or cure) has an important effect on the total number of deaths $D_{\infty} = q R_{\infty}$. 

{\bf Remark 1:\ } It can be shown, using (\ref{14}) and (\ref{15}), that the artificial acceleration of infection (increase of $\beta$, thus decrease of $\kappa$) increases the final number of total deaths.

{\bf Remark 2:\ }The quantity $R_{\infty}$ can also be estimated from the final equilibrium point $I_{\infty}=0$ by solving $I_0 + S_0 + \kappa \log (S_{\infty}/S_0) - S_{\infty}=0$ for $S_{\infty}$ and substituting its value in $R_{\infty}=R_0 - \kappa \log (S_{\infty}/S_0)$.

\paragraph{A sequence of control parameters} Suppose now that a new value $\kappa'$ is achieved at $t=t_f$ for all times in $[t_f , \infty)$. The previously computed quantities undergo a modification

\bea
I_{\rm max} = I(t_f) + S(t_f) + \kappa' (\log (\kappa'/S(t_f)) - 1).
\label{16}
\eea
A piecewise definition of $\kappa$ consists of a set of constants $\{ \kappa_n \}$ at intervals $[ t_n,t_{n+1} ]$. Then, new predictions are obtained for the dreadful $I_{\rm max}$:

\bea
I_{\rm max}(n+1) &=& I(t_n) + S(t_n) + \kappa_n (\log (\kappa_n/S(t_n) -1) \nonumber \\ &=& P- R(t_n) + \kappa_n (\log (\kappa_n/S(t_n) -1).
\label{17}
\eea
Suppose now that $I_n = I(t_n) < I_{\rm max}(n)$ for a sufficiently large $t_n$, i.e. at least one peak has already occurred. The condition for the emergence of a new peak of comparable proportions $I_{\rm max}(n+1)>I_n$ is then 

\bea
I_{\rm max}(n+1) &\sim & I_{\rm max}(n) = I_n + S_n + \kappa_n(\log (\kappa_n/S_n) -1) \nonumber \\
&=& I_{n-1} + S_{n-1} + \kappa_{n-1}(\log (\kappa_{n-1}/S_{n-1}) -1) 
\label{18}
\eea
with $S_n\equiv S(t_n)$ and $S_n + \kappa_n(\log (\kappa_n/S_n) -1) > 0$. Using the total population $P=S_n + I_n + R_n = S_{n-1} + I_{n-1} + R_{n-1}$ in the relation above, the removed differential  $\Delta R = R_n-R_{n-1}$ is shown to satisfy
\bea
\Delta R = \kappa_n \left[ \log \left( \kappa_n / S_n \right) - 1
\right] - \kappa_{n-1} \left[ \log \left( \kappa_{n-1} / S_{n-1}
\right) - 1 \right].
\eea
The new parameter $\kappa_n$ is then given in terms of the old
parameter and the death differential $\Delta D / q$. This quantity
represents a comparison between consecutive regimes with infection
peaks, and $\Delta D >0 $ implies worse scenarios in the evolution. In
order to understand the effect of $\kappa$ on $D$, let us define $x=
\kappa_{n-1} / S_{n-1}$, $x' = \kappa_n / S_n$, such that

\bea
\Delta D / q = S_n x' (\log x'-1) - S_{n-1} x (\log x -1) \equiv S_n
f(x') - S_{n-1} f(x).
\eea
From the dynamical equations, $dS/dt <0$ regardless of $\kappa$, so we
infer $S_{n-1} > S_n $ and we may bound $\Delta D$ from below as

\bea
\Delta D / q > S_n \left[ f(x')- f(x) \right] = S_n \Delta \left[ f(x) \right]
\eea
if $f(x)<0$, and

\bea
\Delta D / q > S_{n-1} \left[ f(x') - f(x) \right] = S_{n-1} \Delta
\left[ f(x) \right]
\eea
if $f(x') < 0$. In the remaining case where both $f(x)>0, f(x')>0$,
then necessarily $x>e$ and $x'>e$, so the infection is either too weak
(large $\kappa$) or we are at the end of the outbreak (small $S$),
which is not interesting for our analysis. In order to estimate
$\Delta D$ in the previous cases, it suffices now to study the sign of
$\Delta f$ in the inequalities above. We have $x, x' \in (0,e)$, thus
$f(x)<0, f(x'<0)$ and $f$ decreases in $(0,1)$, while it increases in
$(1,e)$. Therefore $\Delta f >0$ if $x'<x<1$ or if $1<x<x'<e$. The
first possibility yields $1>\kappa_{n-1}/S_{n-1}>\kappa_n / S_n >
\kappa_n / S_{n-1}$ and so $\kappa_n < \kappa_{n-1} < S_{n-1}$ as
expected, i.e. a smaller $\kappa$ worsens the infection. However the
second pòssibility obtains $eS_n>\kappa_n> S_n \kappa_{n-1}/S_{n-1}$,
which represents a failed quarantine unless $\kappa_n$ grows
beyond $eS_n$, despite being bounded below by $\kappa_{n-1}$.

\subsection{One spatial region with time-dependent control parameters \label{sec:1.2}}

More realistic models of variable social behaviour can be studied. A non-autonomous system with $\kappa = \kappa(\tau)$ can be proposed and solved explicitly in some cases. From the conservation of $P$, we are always left with a two-component system, as $R$ is a redundant function of $S$ and $I$:

\bea
\frac{dS}{d\tau} = -SI, \quad \frac{dI}{d\tau} = \left[ S- \kappa(t) \right] I.
\label{20}
\eea
As we saw before, it is always possible to re-parameterize using one of the SIR variables. For example, let our new time $T$ be the monotonically increasing area under the curve $I$ with time $t$ as independent variable. For constant $\kappa$ this would be tantamount to parameterizing the dynamics with $R$ alone. We have

\bea
T = \int_{0}^{\tau} I d\tau' = \int_{R_0}^{R} \frac{dR}{\kappa(\tau(R))}, \quad \mbox{with} \quad \kappa(T) = \frac{dR}{dT}.
\label{21}
\eea 
Then (\ref{20}) becomes

\bea
\left( \begin{array}{c} S(T)\\ I(T) \end{array}\right) = \left( \begin{array}{c} S(0)\\ I(0) \end{array}  \right) + \int_{0}^{T} dT' \v M \left( \begin{array}{c} S(T')\\ I(T') \end{array}  \right) - \int_{0}^{T} dT' \left( \begin{array}{c} 0\\ \kappa(T') \end{array}  \right)
\label{22}
\eea
with

\bea
\v M = \left( \begin{array}{cc} -1&0\\ +1&0 \end{array}  \right)
\label{23}
\eea
An iteration scheme can be applied to any non-autonomous model for general $\v M(T)$, similar to a Dyson series (physics \cite{sakurai}) or a Volterra series (used by Levy, Wiener and others \cite{volterra1, volterra2}). From (\ref{22}) and the specific case at hand (\ref{23}) we may find the explicit results $S(T), I(T)$:

\bea
S=S_0 e^{-T}, \quad I = I_0 + S_0(1-e^{-T})-\int_{0}^{T} dT' \kappa(T').
\label{24}
\eea
The remaining problem here is to find explicitly the function $T(\tau)$ or its inverse. 

\paragraph{Non-linear, second order equation for lapse} We expect $T(\tau)$ to be governed by a non-autonomous equation, expressing that such kind of systems are not always separable, despite the fact that the existence of solutions is guaranteed by the Picard iterative method. The non-linear, non-autonomous equation for $T(\tau)$ can be obtained by differentiating with respect to $\tau$ the following expression

\bea
\frac{dT}{d\tau} = \frac{1}{\kappa(\tau)} \frac{dR}{d\tau} = I(\tau), 
\label{24b}
\eea
producing thus
\bea
\frac{d^2 T}{d \tau^2} = \frac{dT}{d\tau} \left( S_0 e^{-T} - \kappa(\tau) \right).
\label{25}
\eea
This equation results from the application of the chain rule and the second relation in (\ref{24}). It is easy to see that for constant $\kappa$, (\ref{25}) reduces to a separable equation whose quadrature coincides with (\ref{9}).

\paragraph{Quadrature} Let us solve (\ref{25}) by separation of variables. Here it is important to note that $\kappa(\tau)$ is assumed to be prescribed but, in contrast, $\kappa(T)$ needs the function $T(\tau)$. We might think, however, that the application of some health policy rests on the knowledge of accumulated infections at time $\tau$, which entails the use of $\kappa(T)$ instead of $\kappa(\tau)$ as a specified control parameter. Then, by looking at (\ref{24}) and (\ref{24b}) one has

\bea
d\tau = \frac{dT}{I_0 + S_0(1-e^{-T})-\int_{0}^{T} dT' \kappa(T')}
\label{26}
\eea
and since $\kappa(T)$ is known by assumption, the quadrature 

\bea
\Delta \tau = \int_{0}^{T}  \frac{dT'}{I_0 + S_0(1-e^{-T'})-\int_{0}^{T'} dT'' \kappa(T'')}
\label{27}
\eea
is a solution to (\ref{20}) by direct substitution of the inverse $T(\tau)$ into (\ref{24}). 

\paragraph{Example with an explicit form of the lapse} Suppose a successful policy modelled by $\kappa(T)=\kappa_0 e^T$ is applied. The lapse (\ref{27}) is

\bea
\Delta \tau &=& \frac{\arctan \left[ \frac{A + 2C e^T}{\sqrt{4BC-A^2}} \right]}{\sqrt{4BC-A^2}} \nonumber \\
A &\equiv& I_0 + S_0 - \kappa_0, \quad B \equiv -S_0, \quad C\equiv -\kappa_0
\label{28}
\eea
where the values of $\kappa_0$ are such that the discriminant $A^2-4BC <0$. The function $T(\tau)$ is recovered in terms of the functions $\tan$ and $\log$. For a positive discriminant, $\tanh$ can be used instead. The maximum value of $T$ is reached when $\tau \rightarrow \infty$, and it is determined by the condition $A+Be^{-T}+Ce^{T}=0$.

\begin{figure}[t]
\includegraphics[width=12cm]{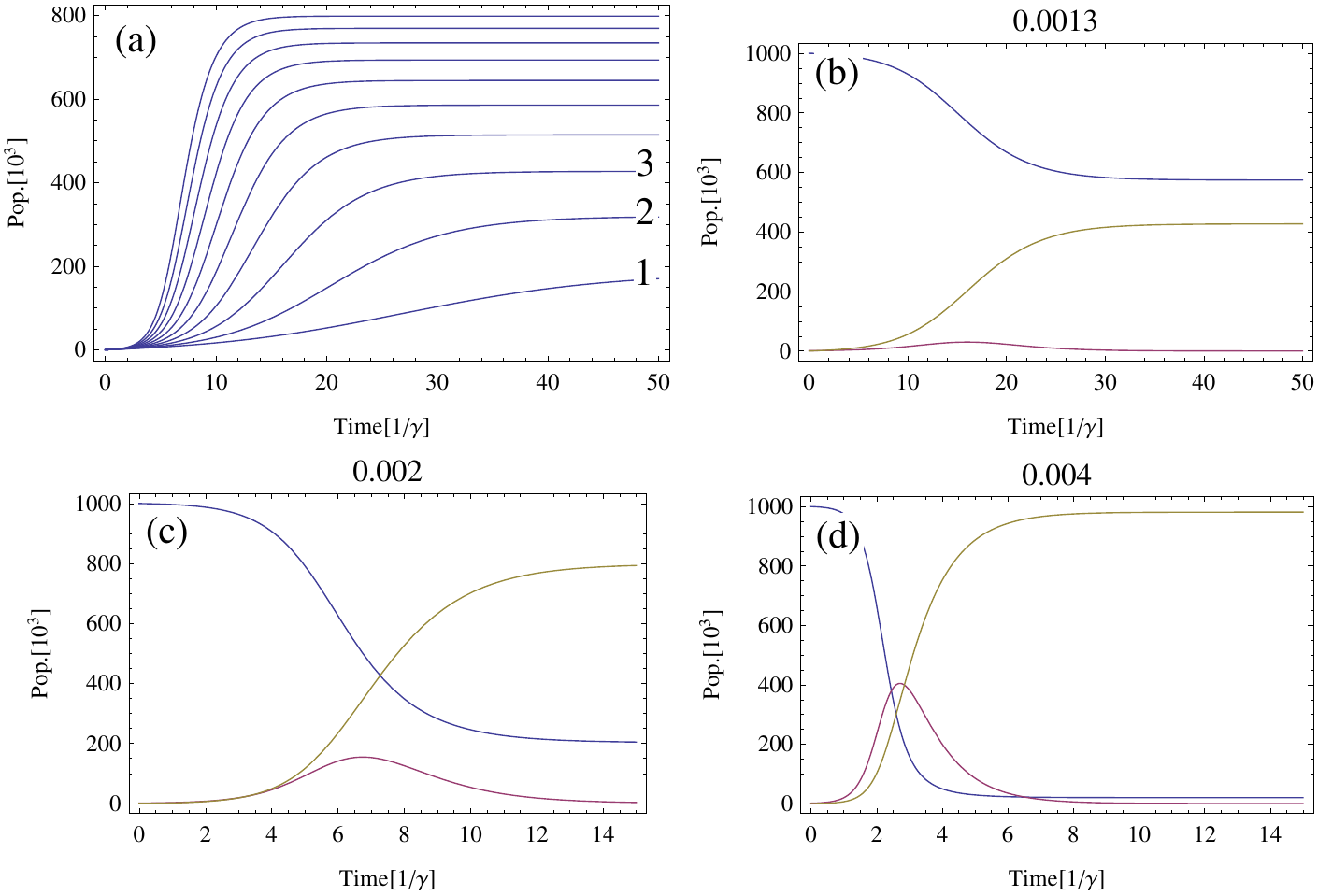}
\caption{\label{fig:1} SIR evolution for various $\kappa$: (a) $R(t)$ increases when $\kappa$ diminishes; curve 1 $\kappa^{-1}=1.1\times 10^{-3}$, curve 2 $\kappa^{-1}=1.2\times 10^{-3}$, curve 3 $\kappa^{-1}=1.3\times 10^{-3}$. (b) Functions $S(t)$ (blue), $I(t)$ (red), and $R(t)$ (gold) with $\kappa^{-1}=1.3 \times 10^{-3}$, producing a small infection curve. (c) Parameter $\kappa^{-1}=2 \times 10^{-3}$ with $R_{\infty}>S_{\infty}$. (d) Parameter $\kappa^{-1}=4 \times 10^{-3}$ with prominent infection curve and $R_{\infty}\approx P$. All curves were generated with initial conditions $S_0=10^6, I_0=10^3, R_0=0$. }
\end{figure}

\begin{figure}[h!]
\includegraphics[width=10cm]{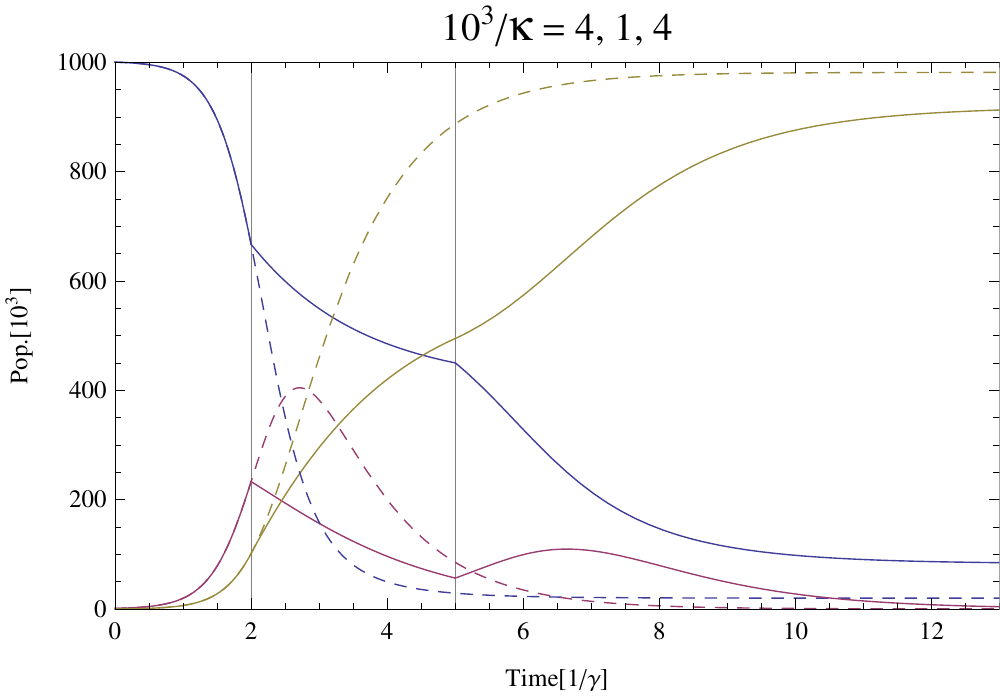}
\caption{\label{fig:2} Evolution with piecewise $\kappa$ parameter, $S(t)$ blue, $I(t)$ red and $R(t)$ gold. For a removal time $\Delta t_r$ of 7 days, the time scale correpsonds to weeks. Between 0 and 2 weeks, $\kappa^{-1}=4\times 10^{-3}$, displaying strong infection. In weeks 2 to 5, a measure is applied $\kappa^{-1}=1\times 10^{-3}$, but from week 5 it is again relaxed to $\kappa^{-1}=4\times 10^{-3}$. A revivial in $I$ is visible around week 7 and the number $R_{\infty} \lessapprox P$.}
\end{figure}

\paragraph{Comparisons} In fig. \ref{fig:1} we show the SIR evolution in weeks. The evolution of the dynamical variables were obtained by numerically integrating Eqs.(\ref{1}) to (\ref{3}) and their behavior is consistent with our analytical expressions. A reasonable recovery time is 7 days, so $1/\gamma = \Delta t_r$ defines our time scale (for COVID-19 it is two weeks \cite{who}). In compliance with \cite{wom}, numbers around 90 to 100 infected per $10^{5}$ inhabitants are reported as world averages; these are used in our simulations: we take as initial conditions  $S_0 = 10^{5}, I_0=10^{2}, R_0=0$ in all cases. We illustrate in panels (b) to (d) the fact that the asymptotic $R_{\infty}, S_{\infty}$ are particularly sensitive to variations of $\kappa$. In fact, the area under the curve $I$ is not constant with $\kappa$, as shown in (a) by plotting $R$. 

In fig. \ref{fig:2} we show the temporal evolution of $S, R$, and $I$ when $\kappa$ also depends on time as a piecewise constant parameter and compare it with the temporal evolution of $S, R$, and $I$ when $\kappa$ is constant throughout the whole epidemic.  Recalling that $\kappa=\frac{\Delta t_i}{\Delta t_r}=\frac{\alpha}{\beta}$, an increase in $\kappa$ may be due to: i) Increase in the removal rate $\alpha$ (i.e, a shorter time to recover) by, for example, improving the medical treatment at early stages or to a quicker time to die, due to perhaps some change in the environmental conditions, or  ii) a decrease in the infection rate ($\beta$), by imposing a quarantine policy, for example. The first two weeks show a quickly evolving outbreak with $\kappa^{-1}=4 \times 10^{-3}$, until a policy is applied, increasing $\kappa$ by a factor of 4. From weeks 2 to 5 the infection curve seems under control, but a relaxed regime starting at week 5 is capable of generating a new peak in $I$. The revival occurs around week 7, and as a consequence, the asymptotics $R_{\infty}$ increase $D_{\infty}$ assuming constant $q$.  Thus,  the effect of increasing $\kappa$ from week 2 week 5, reduces notably the infection peak at the expense of a second milder regrowth and an increase in the duration of the epidemic. At the end the quarantine the removed population decreases by about a half, implying a drastic reduction in the number of deaths at the end of fifth week. However, right after week 5, the removed population picks up again and asymptotically it is only about 10 per cent smaller than the case of a constant $\kappa$ throughout the whole outbreak. Fig. \ref{fig:2} also shows that the effect of a quarentine depends on the initial and final date of its application.

\subsection{Two interacting regions \label{sec:1.3}}

When two regions come into contact by the exchange of susceptible and infected individuals, two species $i=1,2$ of quantities ${S_i,I_i,R_i}$ must be considered. It is reasonable to assume that infections between populations of separate regions are not possible, i.e. $S_1 \rightarrow I_2$ and $S_2 \rightarrow I_1$ are forbidden processes. However $S_1 \leftrightarrow S_2$ and $I_1 \leftrightarrow I_2$ are plausible exchanges. There are two types of migratory trends that can be studied: the depletion of one population into the other and oscillatory dynamics between regions. It is important to note that in the limit of perfect isolation each set of SIR functions evolves as dictated by (\ref{4}). On the other hand, in the limit of negligible infectious rate, the system should decouple into independent migratory subsystems ${S_1,S_2}$, ${I_1,I_2}$ and possibly ${R_1,R_2}$ if removed patients are all recovered and allowed to travel. As before, the total population is conserved here, but the condition now reads $P_1 + P_2 = P$, with $P_i = S_i+ I_i+ R_i$.

The simplest way to take into account these considerations is by introducing an interaction term in our dynamical system

\bea
\frac{dS_1}{d\tau} &=& -S_1 I_1 + F(S_1,S_2), \quad \frac{dS_2}{d\tau} = -S_2 I_2 + G(S_1,S_2), \nonumber \\
\frac{dI_1}{d\tau} &=& (S_1 -\kappa_1) I_1 + J(I_1,I_2), \quad \frac{dI_2}{d\tau} = (S_2 -\kappa_2 )I_2 + K(S_1,S_2), \nonumber \\
\frac{dR_1}{d\tau} &=& \kappa_1 I_1, \quad \frac{dR_2}{d\tau} = \kappa_2 I_2.
\label{29}
\eea
The functions $F, G, J, K$ must comply with conservation of $P$ and, in the absence of infection, the conservation of $S=S_1+S_2$, $I=I_1+I_2$ and $R=R_1+R_2$. Therefore, $F=-G$ and $J=-K$. The change of partial populations is thus $dP_1/d\tau =F+J=-dP_2/\tau$. 

\paragraph{Autonomous vs non-autonomous interaction} Now we distinguish two important cases: Autonomous (time-independent) interaction and non-autonomous (ad hoc) external signal. In the first case we have two subcases, namely a) logistic-type depletion and b) harmonic behaviour.

For a) the interactions are

\bea
F = \lambda (S_1-\sigma) (S_2-\sigma) \Theta(S_1S_2), \quad J = \Lambda (I_1 -\Sigma) (I_2 -\Sigma) \Theta(I_1 I_2),
\label{30}
\eea
with $\lambda$ and $\Lambda$ the intensity parameters given by the migration rate. The quantities $\sigma$ and $\Sigma$ are critical points of the isolated migration model, associated with equilibrium populations. $\Theta$ is the Heaviside step function, which prevents the undesired result $S_i, I_i < 0$ for any $\tau$.

For the subcase b), a harmonic motion can be achieved by employing

\bea
F &=& \omega \,\mbox{Sg}\left[ (S_1-\sigma)(S_2-\sigma) \right] \sqrt{|S_1-\sigma | |S_2-\sigma|} \, \Theta(S_1 S_2), \nonumber \\
J &=& \Omega \,\mbox{Sg}\left[ (I_1-\Sigma)(I_2-\Sigma) \right] \sqrt{|I_1-\Sigma | |I_2-\Sigma|} \, \Theta(I_1 I_2),
\label{31}
\eea
where $\omega$ and $\Omega$ are the frequencies and $\mbox{Sg}$ is the signum function. That this is so, can be inferred from a simple change of variable in typical the harmonic system $\dot x= \omega y, \dot y = -\omega x$, where $x^2+y^2$ is constant. Indeed, $X=x^2$ and $Y=y^2$ leads to $\dot X = 2 \omega \sqrt{|XY|} = - \dot Y$.

When the interactions are modelled by external ``signals", a subcase b) can also be considered. Oscillations can be treated with

\bea
F= \frac{a}{\omega} \cos (\omega \Delta\tau), \quad J= \frac{A}{\Omega} \cos (\Omega \Delta\tau),
\label{32}
\eea
where $a,A$ are oscillation amplitudes.

{\bf Remark 3.\ } External signals such as (\ref{32}) or damped signals with factors $e^{-\Gamma \Delta \tau}$ are useful for reproducing revivals. Bimodal curves for $I_i$ indicate cycles of infection parameterized by the frequency $\Omega$.

{\bf Remark 4.\ } Interacting models such as (\ref{31}) display rigidity for critical frequencies. A variety of numerical methods based on Runge-Kutta algorithms diverge at finite times.

{\bf Remark 5.\ } The total population of two interacting regions does not behave as a full SIR, even in the absence of migration or exchange. Under the change of variables $S=S_1+S_2$, $I=I_1+I_2$, $R=R_1+R_2$, $s=S_1-S_2$, $i=I_1-I_2$, $r=R_1-R_2$, $K=(\kappa_1+\kappa_2)/2$, $\kappa=(\kappa_1-\kappa_2)/2$ the system (\ref{29}) picks effective interactions. The large system is 

\bea
\frac{dS}{d\tau} = -\frac{1}{2} \left[ SI + si \right], \quad 
\frac{dI}{d\tau} = \frac{1}{2} \left[ (S-K)I + (s-\kappa)i \right], \quad \frac{dR}{d\tau} = K I + \kappa i,
\label{33large}
\eea
and the small subsystem is

\bea
\frac{ds}{d\tau} = -\frac{1}{2} \left[ Si + sI \right]+2F, \quad 
\frac{di}{d\tau} = \frac{1}{2} \left[ (S-K)i + (s-\kappa)I \right] + 2J, \quad \frac{dr}{d\tau} = \kappa I + K i. \nonumber \\
\label{33small}
\eea
Systems comprising two or more regions are, in general, non-separable. 

\begin{figure}[t]
\includegraphics[width=12cm]{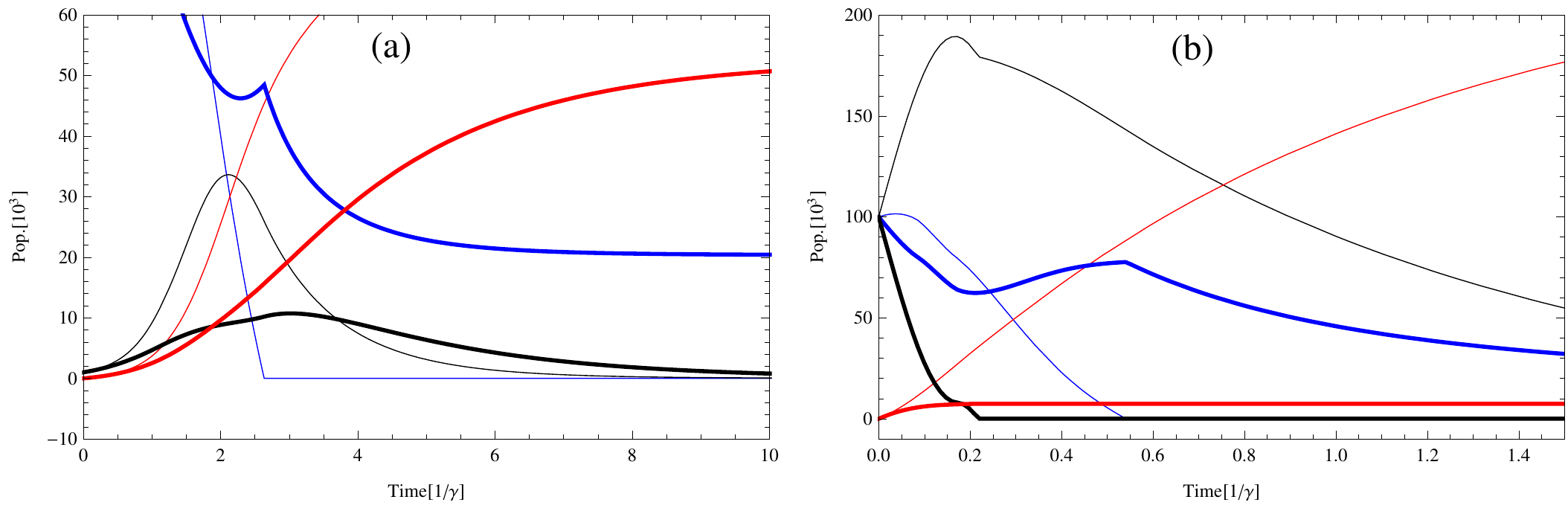}
\caption{\label{fig:3} Autonomous harmonic interaction between two regions. Curves: $S_1$ blue, $I_1$ black, $R_1$ red, $S_2$ blue (thick), $I_2$ black (thick), $R_2$ red (thick). Parameters: for (a) $\alpha = 0.03 = \beta, \gamma = 1 = \delta,
\omega = 1, \Omega = 0.1, \sigma = 0.8\times 10^{5}, \Sigma = 0; S_1(0) = 10^{5} = S_2(0), I_1(0) = 
 10^{2} = I_2(0), R_1(0) = 0 = R_2(0)$, for (b) $\alpha = 0.01 = \beta, \gamma = 1= \delta,
\omega = 9, \Omega = 9, \sigma = 0.8\times 10^{5} = \Sigma, S_1(0) = 10^{5} = S_2(0) = I_1(0) = I_2(0), R_1(0) = 0 = R_2(0)$. }
\end{figure}

\begin{figure}[h!]
\begin{center}
\includegraphics[width=8cm]{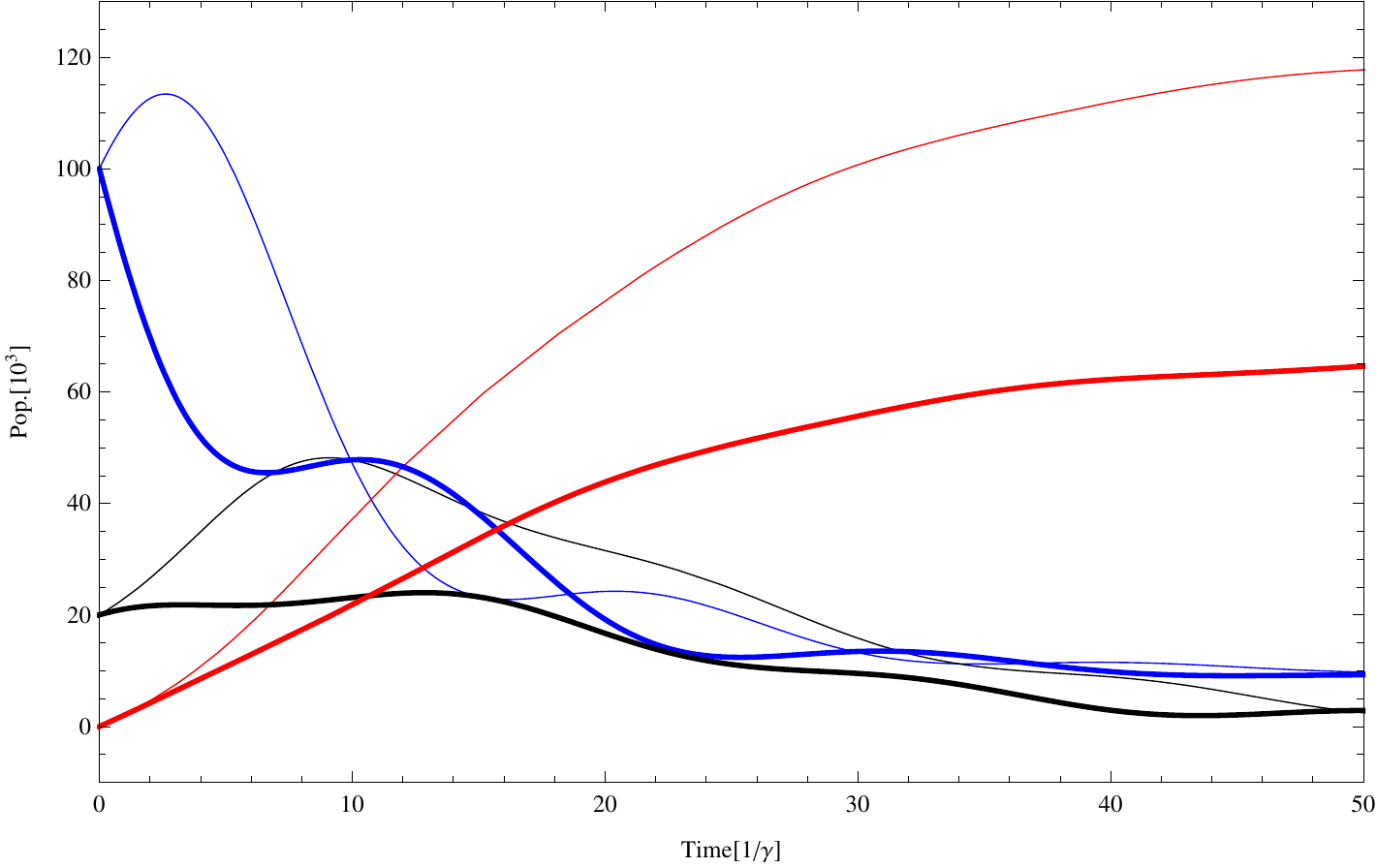}
\end{center}
\caption{\label{fig:4} Oscillatory migration with non-autonomous harmonic term. Curves: $S_1$ blue, $I_1$ black, $R_1$ red, $S_2$ blue (thick), $I_2$ black (thick), $R_2$ red (thick). Parameters: for (a) $\alpha = 0.002 = \beta, \gamma = 0.1 =\delta, \Gamma_1 = 0.1, \Gamma_2 = 0.01, \omega = 1/3 = \Omega, a = 40 \omega/3, A = 2 \Omega/3, 
I_1(0) = 2\times 10^{3}=I_2(0), S_1(0) = 10^{5}=S_2(0), R_1(0) =0=R_2(0)$. The damping for population $i=1,2$ is given by $\Gamma_i$.}
\end{figure}

\begin{figure}[h]
\includegraphics[width=12cm]{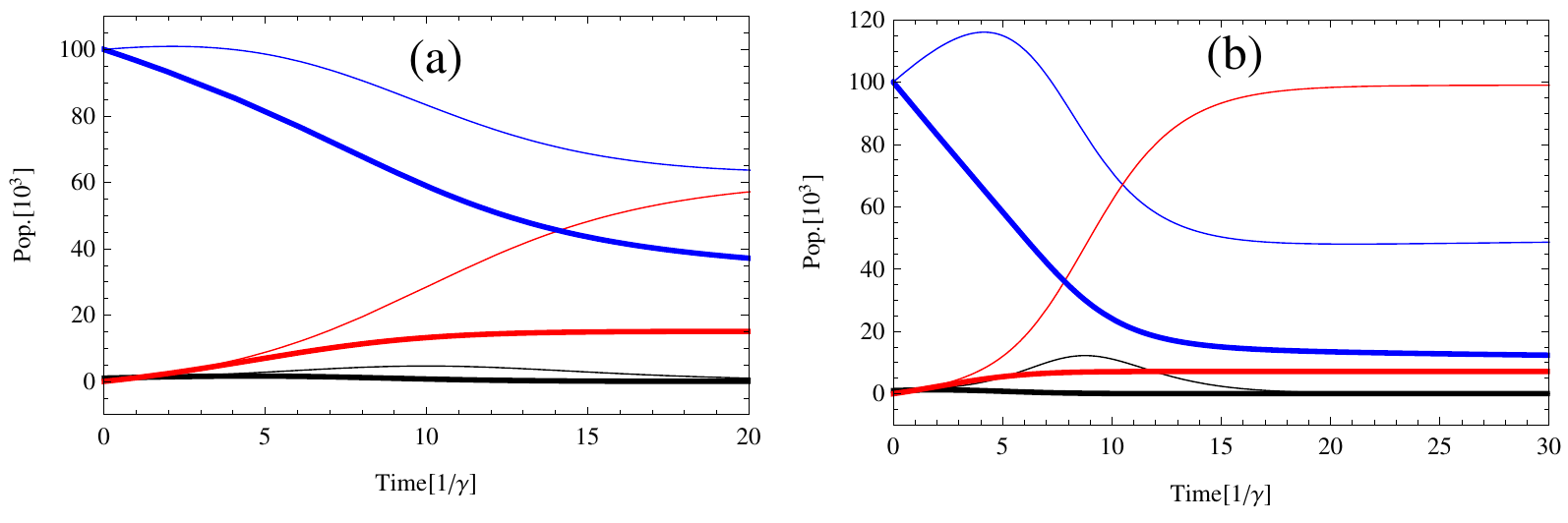}
\caption{\label{fig:5} Migratory depletion in a system of two regions. Curves: $S_1$ blue, $I_1$ black, $R_1$ red, $S_2$ blue (thick), $I_2$ black (thick), $R_2$ red (thick). Parameters: for (a) $\alpha = 1.2 \times 10^{-2}= \beta, \gamma = 1= \delta, \lambda = 2.5 \times 10^{-4}, \Lambda=0, \sigma = 10, \Sigma=0, I_1(0) = 10^{3}, S_1(0) = 10^{5}, R_1(0) = 0, 
I_2(0) = 10^3, S_2(0) = 10^{5}, R_2(0) = 0$ and for (b) $\alpha = 0.0120 = \beta, \gamma = 1 = \delta, \lambda =
9\times 10^{-4}, \Lambda=0, \sigma = 10^{4}, \Sigma = 0$ and the same initial conditions as in (a). }
\end{figure}

\paragraph{Results} For regions interacting with autonomous harmonic terms such as (\ref{31}), we can show that the infection peak can display anomalous behaviour in different scenarios. The typical outbreak with initial conditions $S_1(0) = 10^{5} = S_2(0), I_1(0) = 10^{2} = I_2(0), R_1(0) = 0 = R_2(0)$ is shown in Fig. \ref{fig:3} panel (a). There we see a shoulder in the infection curve (black thick line) and an anomalous revival in the second susceptible population (blue thick line) whereas in panel (b) another set of initial conditions $S_1(0) = 100 = S_2(0) = I_1(0) = I_2(0), R_1(0) = 0 = R_2(0)$ describes the situation once the pandemic has been run until $S$ and $I$ are comparable. In such a case we also see an infection peak with a shoulder (black thin line) for the first population, but a drastically decreasing $I_2(t)$ (black, thick line). A more conspicuous oscillatory effect can be achieved by non-autonomous terms in (\ref{32}): The results in fig. \ref{fig:4} show an infection curve with notorious damped oscillations for parameters $I_1(0) = 2\times 10^{3}=I_2(0), S_1(0) = 10^{5}=S_2(0), R_1(0) =0=R_2(0)$. The frequencies are $\omega=\Omega=1/3$ (in units of $\gamma$). For the second type population, an infection curve with multiple ``humps" in thick black is visible. Lastly, the results of a migratory depletion model in (\ref{30}) are displayed in fig. \ref{fig:5}. The difference between panels (a) and (b) comes from a larger migration rate for (b), albeit the use of similar initial conditions. As a result, the infection curve for population 1 (black thin line) has a larger peak in (b), in compliance with a migratory trend from region 2 to 1.

\subsection{Total distributions \label{sec:1.4}}

Our previous remark establishes that separate regions produce a non-trivial joint effect even when they are isolated. Also, in the presence of migration, the evolution of total SIR populations of a country or large region unfolds differently as compared to non-interacting regions.
\paragraph{Uncorrelated case} Let us suppose that during an outbreak a set of disconnected regions  evolve under different, uncorrelated, conditions. The resulting individual curves will have similar shapes but dfferent time shifts and peak intensities. For example, the infection processes could have different starting points, different initial conditions and varying values of $\kappa$. 
It is then natural to consider a full distribution made of individual populations with a weight factor $W$ given by the local population fraction (e.g. France in the world: $(6.699 \times 10^7)/(7.8 \times 10^9)= 8.588 \times 10^{-3}$). Using a monotonic parameterization of such weight $W(a)$, $a \in [0,1]$ one has

\bea
S(t) = \int_{0}^{1} da W(a)S_a(t), \quad I(t) = \int_{0}^{1} da W(a)I_a(t), \quad R(t) = \int_{0}^{1} da W(a)R_a(t).\nonumber \\ 
\label{34}
\eea
The integration has a smoothing effect in global measurable quantities. For instance, in a strict quarantine, the resulting curve for world populations in major scale events (see COVID-19 in \cite{wom}) has a regular behaviour in limited time windows, as if modelled by a global SIR with effective $\kappa$. However, each region has a distinct behaviour governed by a local $\kappa$ and entirely different sets of curves. With this, we would like to stress that the practice of performing {\it a posteriori\ }curve fitting does not necessarily validate the correctness of dynamical models locally.

\paragraph{Interacting case} The set of equations for fully interacting regions become

\bea
\frac{d S_a}{d \tau} &=& - S_a I_a + \int_{0}^{1} da' F_{a,a'}(S_a,S_a'), \nonumber \\
\frac{d I_a}{d \tau} &=& (S_a - \kappa_a) I_a + \int_{0}^{1} da' J_{a,a'}(I_a,I_a'), \nonumber \\
\frac{d R_a}{d \tau} &=& \kappa_a I_a.
\label{35}
\eea
where the consistency requirements of conserved populations are now

\bea
\frac{dP}{d\tau} &=&0 \, = \,\frac{d}{d\tau} \int_{0}^{1} da P_a = \int_{0}^{1} \int_{0}^{1} da da' \left[ F_{a,a'}(S_a,S_a') + J_{a,a'}(I_a,I_a') \right], \nonumber \\
\lim_{ \beta \to 0} \frac{d S}{d t} &=& 0 \, = \, \int_{0}^{1} \int_{0}^{1} da da' F_{a,a'}(S_a,S_a'), \nonumber \\
\lim_{ \beta \to 0} \frac{d I}{d t} &=& 0 \, = \, \int_{0}^{1} \int_{0}^{1} da da' I_{a,a'}(S_a,S_a').
\label{36}
\eea
Note that the second and third relations contain $t$ and not $\tau$ in the limit.
Once more we recall here that the absence of infection reduces to a model of circulation with conserved quantities SIR. More detail can be introduced if $a$ is replaced by coordiantes $\v r$ and $F_{a,a'}, I_{a,a'}$ are replaced by generalized distributions containing $\nabla_{\v r} \delta^{(3)}(\v r - \v r')$, $\partial^{i}_{\v r} \partial^{j}_{\v r} \delta^{(3)}(\v r - \v r')$, and so on, which give rise, upon integration, to local operators such as $\nabla, \nabla^2$ etc. From here we may obtain as special cases the diffusion equations and continuity conditions in the form of Fick's law.

\section{Reduction to logistic models with applications to current outbreaks \label{sec:2}}

The logistic model  and its generalizations  are useful in the description of growth for many applications, including biology. Phenomenological uses were originally proposed by Richards \cite{richards1959}, based on historical discussions by Gompertz \cite{gompertz} and previous work by Von Bertalanffy \cite{vonbert1, vonbert2}. Such logistic functions can be viewed as a particular reduction of SIR models to a single variable, i.e. the accumulated population $A$, which is the time integral of $I$ under the condition $P=S+I$. This corresponds to the situation ``once infected, always infected". The reduction gives rise to an integrable model by quadrature. Moreover, it allows to integrate the flow in terms of elementary functions for a variety of non-linear systems that generalize the autonomous logistic equation. One of our aims here is to see how capable is the model, with constant parameters throughout the whole outbreak, to give reasonably good fits to the data and to  express the model's sensitivity to parameter variations and data fluctuation; particular attention shall be paid to the total number $D$ and its fluctuations. This can be easily done with analytic solutions at our disposal. 

Richards' model for the accumulated number of infected population $A$ is defined by the flow 

\bea
\frac{d A}{dt} =F(A;r,\alpha,K)\equiv r A \left[ 1 - \left( A/ K \right)^a \right]
\label{RichFlow}
\eea
where the accumulated population $A$ is given in terms of new infections $I_{\rm new}$
\bea
A(t) \equiv \int_{t_0}^t dt' \quad I_{\rm new}(t') .
\label{38}
\eea
The parameters are $r$: modulation of the infection rate, $K=A_{\infty}$: final accumulated  count of infected population , and $a$ plays the role of recovery coefficient; the larger $a$ is the faster the infection ends.  The parameter $a$ also determines the asymmetry of the curve of the daily new infections: -- e.g. $a=1$ corresponds to a symmetric curve around $A=K/2$ -- with fixed points $A=0, A=K$. The allowed values of the parameters are determined by the existence of such fixed points; we have $r>0, a>0$ and $r<0, -1<a<0$, satisfying the condition $ra > 0$. In particular, $a \in (-1,0)$ generalizes the standard logistic map, where $a=1$.

\subsection{Exact solutions, constant parameters \label{sec:2.1}}

Assuming constancy of the  paramters $r,a,K$,  the autonomous system has solutions of the form

\bea
r \Delta t = \int_{A_0}^{A} \frac{dA}{A(1-(A/K)^a)},
\label{39}
\eea
which translate into
\bea
A(t)= \frac{e^{r\Delta t} A_0 K}{\left[ K^a + \left( e^{a r \Delta t} -1 \right) A_0^a \right]^{1/a}}.
\label{SolA(t)}
\eea
{\it{Robustness, data fluctuations}} It is straightforward to compute the fluctuations in the predictions of final accumulated numbers of infected patients. The problem of interest is to analyze the error in forecasting due to bad data at a given time. Assuming that such statistical error at time $\Delta \tau$ is $\delta A$, then the constant $K$ changes to $K + \delta K$. To see this, one simply solves for $K$ in (\ref{SolA(t)}) as a function of $A$ and $\tau$ and proceeds with the partial variation of $A$ as independent variable, according to

\bea
K(A+\delta A; \tau, a, A_0,...) = K(A; \tau, a, A_0,...) + \frac{\partial K(A; \tau, a, A_0,...)}{\partial A} \delta A + ...
\label{42}
\eea
Since $A \rightarrow K$ as $\tau \rightarrow \infty$, we expect $\delta K \rightarrow \delta A_{\infty}$, as can be verified explicitly. Small errors in data produce then the variation

\bea
 \delta K=\delta A \frac{A_0(e^{-a\Delta \tau}-1)^{1/a}}{A_0^a-A^a e^{-a\Delta \tau}} \frac{A_0^a}{A_0^a-A^a e^{-a\Delta \tau}}=\delta A \frac{K}{A}\frac{A_0^a}{A_0^a-A^a e^{-a\Delta \tau}},
\label{43}
\eea
where the first fraction has been identified as the function $K(A,A_0)$ obtained from (\ref{SolA(t)}).
Expression (\ref{43}) supports the concept that final predictions have different sensitivities at different times. For example, the variation on $K$ due to a variation (bad data)  of $A$ at $\delta \tau$ very large (that is, near the end of the pandemic) is easily seen to be $\delta K=K\frac{\delta A_{\infty}}{A_\infty}=\delta A_\infty$, a negligible amount. In contrast,  if the variation occurs at early times ($\Delta \tau \approx 0$) then the impact on the prediction for $K$ is $\delta K=\delta A_0 \frac{K}{A_0}$, which can be very large even if $\delta A_0$ is small. Since this expression for $\delta K$ is a monotonously decreasing function of time, the errors due to bad data at some time $t$ gradually become less important as $t$ increases.

\subsection{Examples of curve fitting and forecasting with Richards model \label{sec:2.2}}

Here we try Richards model to fit Mexico's COVID-19 data \cite{cibrian2020} for forecasting and as an illustration of the sensitivity to small variations in data, obtained at various times. See also \cite{mexforecast}. First we obtained analytical expressions for relevant quantities, such as time of peak infections, number of daily new infections at its peak, etc.

Integration of  (\ref{RichFlow})  with initial condition $A_0$  gives

\begin{equation}
ln(B_0/B_t)= ar(t-t_0),\label{gensol}
\end{equation}
where
\begin{equation}
B(t)= (K^a-A(t)^a)/A(t) ^a.
\end{equation}\label{Boft}

%Solving (\ref{gensol}) for $I(t)$ gives
%\begin{equation}
%I(t)= \frac{K}{(1+((K/I_0)^a-1) exp(-r a(t-t_0))^{1/a}}\label{Csolution}
%\end{equation}
%and solving (\ref{gensol}) for $t$ gives
%\begin{equation}
%t=t_0 +\frac{1}{ar}ln(b_0/b(t)).
%\end{equation}

Solution $A(t)=$(\ref{SolA(t)}) follows from  (\ref{gensol}). It is interesting to determine the time $t_p$ at which the speed of infection reaches its peak.This occurs when 
$F(A;r,a,K)$ reaches its maximum value, i.e, when $\frac{\partial F}{\partial A}=0$:
\begin{equation}
A_p=A(t_p)=\frac{K}{(1+a)^{1/a}}.\label{Cp(a)}
\end{equation}
It follows  that
\begin{equation}
B(t=t_p)\equiv B_{t_p}=a.
\end{equation}
Hence, 
\begin{equation}
t_p=t_0+ \frac{1}{ar}ln(B_0/a).\label{tp(a)}
\end{equation}

Note that the term $(1+a)^{-1/a}$ in eq.(\ref{Cp(a)}) is a monotonically increasing fiunction of $a$, growing $0$ at $a=-1$,  reaching $1/e$  at $a=0$ and approaching $1$ at $a\rightarrow \infty$ . The quantity $(1+a)^{-1/a}$ can be roughly approximated by the straight line $\frac{1}{e}+\frac{e-2}{2e}a$. So  $A_p$  grows roughly linear with  $a$. Note that (\ref{Cp(a)}) limits the range of $a$ to $[-1,\infty)$  and, complementary,  (\ref{tp(a)}) indicates that negative values of $a$ are allowed provided $r$ is negative simultaneously with $a$. Also note that there is no problem with $a$ being negative in the expression for $t_p=(\ref{tp(a)})$ since $B_0$ automatically becomes negative as $a$ becomes negative \\

It is also useful to determine the temporal evolution of {\it  daily new cases}. In particular, we wish to know, for a given set of values of  $(a, r ,K)$ when and what is the highest number of daily new cases. This information can be obtained numerically using the solution (\ref{RichFlow}) and computing  $\Delta A(t)= A(t)-A(t-1)$.  It can also be obtained by noting that $\Delta A(t)=\frac{\Delta A(t)}{\Delta t}\approx \dot A = F(A(t);a,r,K) $, where $\Delta t=1$. This approximation is excellent since the function $A(t)$ is very smooth for all $t$.  For the particular  time $t=t_p$ the value $F(A(t_p);a,r,K)=rA_p(1-(A_p/K)^a)$ gives  the maximum increase in daily new cases and can readily be obtained using (\ref{Cp(a)}):
\begin{equation}
\Delta A(t_p)=raA_p/(1+a)=rKa/(1+a)^{1+1/a} \approx rKa/4.\label{Deltacp}
\end{equation}

The fitting parameters $(a,r, K)$  of solution (\ref{SolA(t)}) depend on the choice of precision of the fitting and on the number of data used to perform the fitting. Fitting precision is controlled by the maximum number of  iterations $M$ given as an input in a subroutine that incorporates the method of least squares. Increasing $M$ improves the fitting up to a certain value $M$, after which the values of the parameters remain almost unchanged, according to a prescribed tolerance. Numerically we found $M=1150$.
Since the data are not very reliable due to various factors, such as the lack of actual laboratory testing in all suspected cases and the delay of gathering daily data, it is not indispensable to make as precise a fitting as possible. In this context, we may regard various different fitting precisions as possible different epidemic scenarios. 

\begin{table}[t]
 \centering
 %\large
  \begin{tabular}{|c|c|c|c|c|c|c|c|}  \hline    \hline
Case&~$(n_i,n_f)$           & ~$a$~       & ~$r$~  & ~$K$~  &  $(A_p,\Delta A_p)$ &~$t_p$~&~$\sigma$~\\ \hline
I$_+$& ~$(1,118)$ ~   &    0.0002588    &    77.3681     &633787       & (233187, 4669) &~ 128(July 4) ~&  ~1060~       \\ \hline
%$k_{B,1}$               &  1.67715      & 4.43494        & 4.65501          & 2.08837   &  ~~0.0207~~        \\ \hline
I$_-$ &  ~  $(1,118 )$ ~         & -0.08728      &     -0.1460    & 1313370         &  (461274, 6441)   &  ~163(Aug 8 ) &  386       \\ \hline
 II$_+$&~ $(25,118)$ ~               & 0.000279     & 68.1746    & 699919        & (257522, 4900) &  ~132 (July 8)~&  ~927      \\ \hline
II$_-$&~ $(25,118)$ ~      &-0.12713      &  -0.0866   & 1614580       & (554078, 6986)   &  ~175 (Aug 20)&  ~419~~   \\ \hline
III$_+$&~ $(1,134)$ ~      &0.000185     &   103.54   & 742090       & (273025, 5217)   &  ~134 (July 10)&  ~~1856~~ \\ \hline
III$_-$&~ $(1,134)$ ~      &-0.09335      &  -0.131129   & 1419160       & (496717, 6706)   &  ~167 (July 12)&  ~~589~~\\ \hline
IV$_+$&~ $(25,134)$ ~      &0.000338      &   53.8848   & 815251       & (299965, 5462)   &  ~138 (July 14)&  ~~1562~~ \\ \hline
IV$_-$&~ $(25,134)$ ~      &-0.127318     & -0.086317   & 1620450       & (556030, 7002)   &  ~175 (Aug 20)&  ~~636~~ \\ \hline

%8    & 1150/118      & 0.000124     & 161.67   & 634306 & ~~0.0200~~       \\ \hline 
%$2/l(k_{1}^*)$    & 0.4112      & 0.0355      & 0.0343    & 0.0372          \\ \hline \hline
 % B                        &  0.21231      & -31.2604        & 4.57223        & -2.99219        \\ \hline
\end{tabular}
\caption{Fitting parameters and predictions including negative values of $a$ and $r$. Here and elsewhere $n_i (n_f)$ is the initial (final) day of the fitting data}.
\label{Table1}
\end{table}

 In what follows we shall fit the solution (\ref{SolA(t)}) of  Richards model to the data of cumulative  number of infected people of COVID-19 in Mexico (data gathered from \cite{wom} www.worldometers.info/coronavirus). The first case of infection was detected on February 28, 2020, the day number 1. Table \ref{Table1} shows the values of  parameters $(a,r,K)$ and the predictions $t_p, A_p$, and  $\Delta Ap$,  obtained from 8 representative fittings for two sets of data; one from day 1  till day 118 (June 24, 2020); the second from day 1 till day 134 (July 10, 2020).
 
  The first four fittings  ($I_+, I_-, II_+$, $II_-$) used the data of the first set and the last four cases ($III_+, III_-, IV_+$, $IV_-$) used the second set of data.  The sub index in each case refers to the sign of the parameter $a$ (both signs are allowed, see discussion below (\ref{tp(a)})). Further, fitting $I_+$ differs from fitting $II_ +$ in that the latter does not include the first 25 days of the data.  The same distinction is true for the pairs $(II_+, IV_+)$. This was done in order to test if predictions could be improved by resting importance to the initial data and to test how much the fitting parameters change with time. Comparing  the values of $K, A_p, \Delta A$,  $ t_p$, and $t_f$ for the pair ($I_+, II_+$) shows that, indeed, disregarding the initial 25 data points yields somewhat larger values  of  the number of total accumulated infections and the duration of  the epidemic. As we shall see below, the larger values of  $K,t_p, A_p$  makes the extrapolation of the fitting $II_+$ come closer to the data values for a few days after the end of the fitting, the day 118. The root mean square deviation $\sigma$, a measure of how much the fit deviates from the data,  is appreciable smaller for the case $II_+$ than  for the case $I_+$, indicating that case $II_+$ is a better fitting. This improvement on the fitting correlates with an improvement on the short term prediction.  The same occurs between cases $III_+$ and $IV_+$; this time with 134 as the final day of fitting; here $\sigma$ changes from 1856 to 1562  as the  first 25 days are ignored.
  
\begin{figure}[h!]
\centerline{\includegraphics[width=0.95\columnwidth]{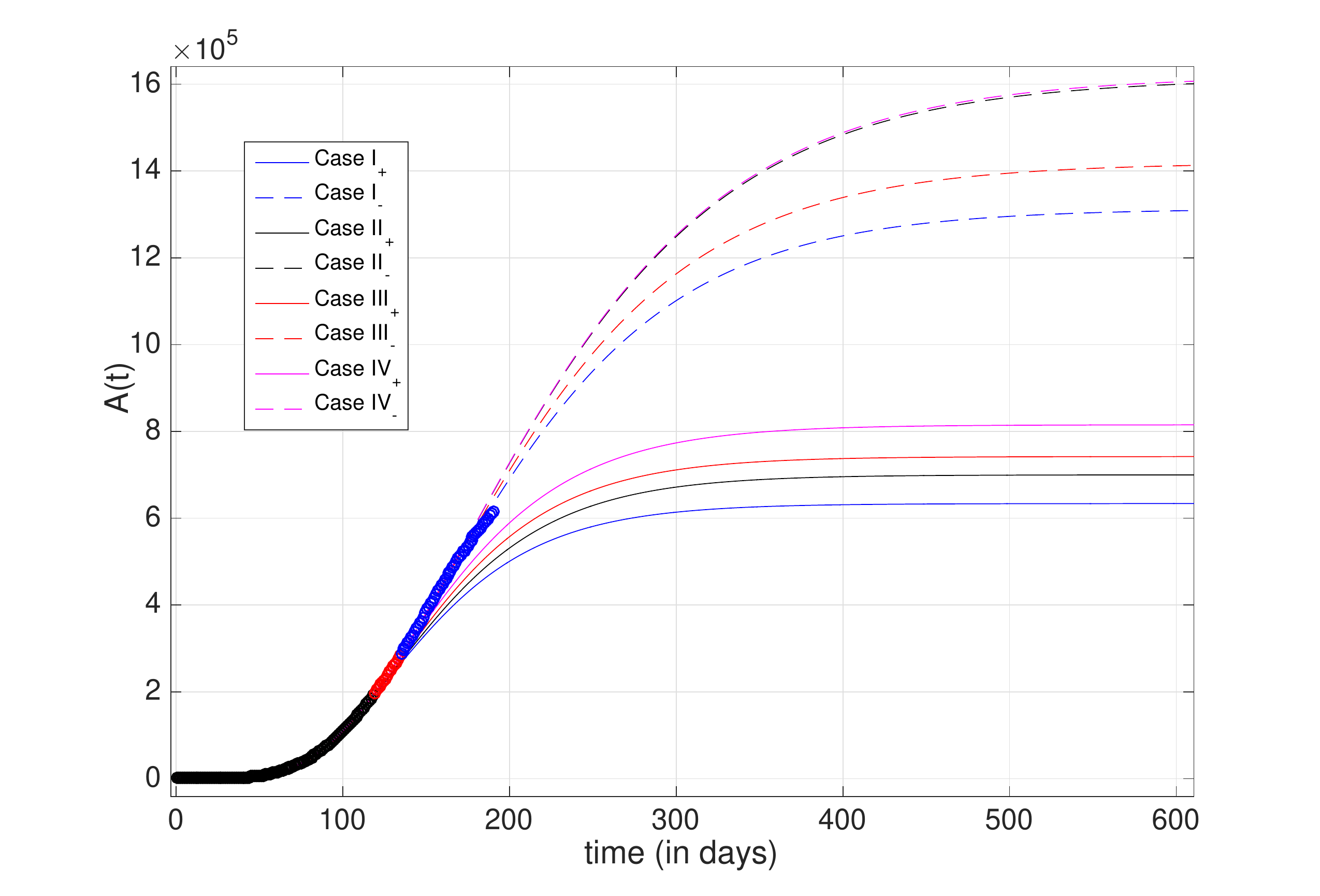}}
\caption{Accumulated number of infections $A(t)$  for 8 cases listed in Table 1 versus data up to day 187 (September 1st, 2020).}
\label{fig 6}
\end{figure}

  In Figure \ref{fig 6} we plot  $A(t)= (\ref{Cp(a)})$ for the 8 cases listed in Table 1 together with the actual data. Figure 7 is a zoom of Fig. 6. The above mentioned  cases  correspond to the four lowest curves, $I_+$ (blue solid line), $II_+$ (black solid line), $III_+$ (red solid line) and $IV_+$ (magenta solid line). These illustrate that even though cutting off the first 25 day of fitting does improve the predictions it does it only for a week or so after the fitting final date. 
  
It is evident that all 8  fittings give similar results for the first stages of the pandemia but their asymptotic behavior is quite different, fanning out into two sets. Specifically the set  that diverts very soon from the data corresponds to those fittings with positive values of $a$ and $r$. The second set of   fittings have all  negative values of $a$ (simultaneously, negative values of $r$)  is able to predict the data correctly and for a much longer time. 

The reason negative values yield better predictions is that the data for the daily new cases $\Delta A$ is far from being symmetric about its peak value $A_p$. It is symmetric when the peak $A_p$ is $K/2$.  Richards model allows for the asymmetry through the parameter $a$.  As  Eq.(\ref{Cp(a)}) shows, the flow $F(A;r,\alpha, K)=(\ref{RichFlow})$ is symmetric only  for $a=1$  ( the standard logistic flow). As $a$ becomes less than 1, the peak moves to the left of $A= K/2$ and as $a\rightarrow 0^+$, $A_p \rightarrow K/e$. So, the solution (\ref{SolA(t)}) would be unable to provide good fittings with $a>0$ if the data demands that $A_p$ be smaller or equal to  $K/e$. This argument is consistent with the ratios computed for  all  the fittings we have done. In particular, the ratios $A_p/K$  for the cases $I_+,II_+, III_+$ and $IV_+$ are, respectively, $0.36793, 0.36793, 0.36791,$ and $3.6794$; these are barely larger than $1/e= 0.36787...$. The other 4 cases have negative $a$ (and $r$) and the ratios are little smaller than 1/e; namely, $0.3512, 0.34317, 0.35791$, and 
$0.34313$ for  $I_-,II_-, III_-$ and $IV_-$, respectively.
Note that as $a$ becomes negative, $r$ becomes negative also in order to preserve the  concavity of the flow function (\ref{RichFlow})
 
\begin{figure}[t]
\centerline{\includegraphics[width=0.95\columnwidth]{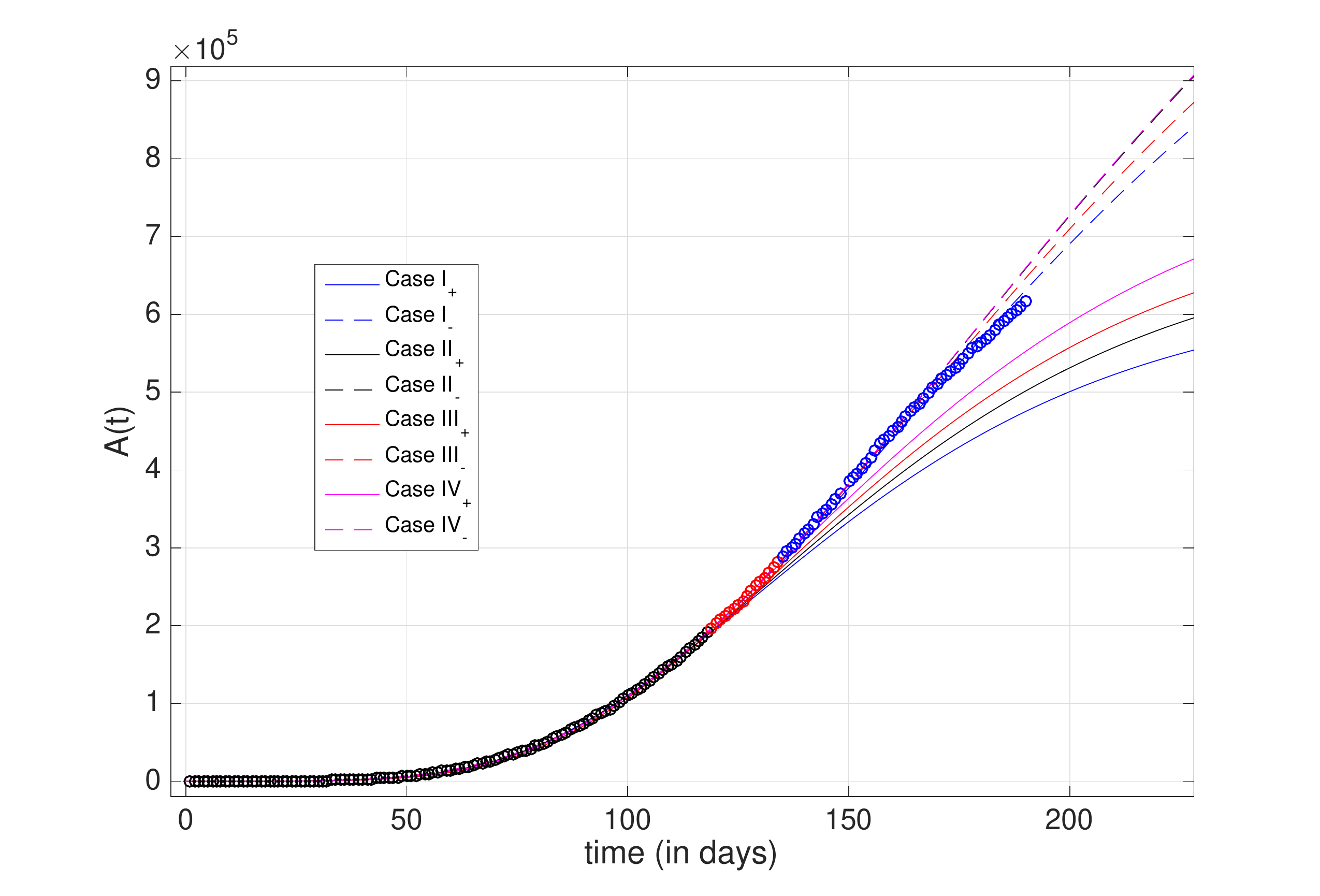}}
\caption{Accumulated infections. Zoom of Fig.6.}
\label{fig 7}
\end{figure}

In conclusion, Figs. \ref{fig 6} and \ref{fig 7}  show that the fittings  with negative  values of $a$ and $r$ render not only a better fitting but their extrapolations after the final fitting date are able to predict quite well the data for about 2 months. Still, there is no certainty that the predictions will continue to agree with data beyond a few days after the day 188 ( September 2, 2020) but one may venture to forecast that the actual evolution of $A(t)$ will follow a curve close to the fitting $I_-$ with a total number of infected population of about 1.1-1.3 million people, or close to 1 percent of the total  population in Mexico.

\subsection{Discretization and Bifurcation \label{sec:2.3}}

For any given country or region the generation of data for epidemics is updated daily, so it is interesting to see what differences may occur for the discretized version of  Richards model.
Rescaling the cumulative number of infections $A(t)$  to the total accumulated number of infections $K:  A\rightarrow x\equiv A/K$ the discretized version of (\ref{RichFlow}) is the mapping

\begin{equation}
x_{n+1}=G ((x_n; \alpha, r)=(1+r)x_n -rx_n^{a+1},
\end{equation}

where we have taken the  constant lapse time $\Delta t$ between gathering of data  as our  unit of time.  We can further simplify the flow  (\ref{RichFlow}) , by  rescaling the time by the infection coefficient $r$, $t\rightarrow \tau \equiv r\times t$, leading to
\begin{equation}
\frac{d y}{d\tau}= y(1-y^\alpha).
\end{equation}
Note that  we have changed the notation for the scaled number of infections $x=I/K$ to $y=I/K$ since $x( t)$ is in general not equal to $x(r\times t)\equiv y(\tau)$.   
Its corresponding mapping is
\begin{equation}
y_{n+1}= H_\alpha(y_n) \equiv 2y_n-y_n^{\alpha+1}.
\label{simplemap}
\end{equation}
Here, our unit of lapse time is $\Delta \tau= r \Delta t$. We now  list the basic properties of this simpler  homeomorphism

\begin{itemize}

\item[{\it i)\ }] {\it Zeroes of $H_\alpha(y)$.\ }
\begin{equation}
H_\alpha(y)=0\quad \mbox{at} \quad y=0 \quad  \mbox{and} \quad  y \equiv y_e=2 ^{1/\alpha}.
\end{equation}
 
\item[{\it ii)\ }] {\it There are only two period-one fixed points:\ } 
\begin{equation}
H_\alpha(y)=y \quad \mbox{at}  \quad y=0 \quad\mbox{and}\quad y=y_p=1\quad \forall \quad \alpha
\end{equation}
 
\item[{\it iii)\ }] {\it $H_\alpha$ has only one critical point $y_c$.\ } $\frac{dH_\alpha}{dy}=0$ for

\begin{equation}
y=y_c\equiv 2^{1/\alpha}/(1+a)^{1/\alpha} 
\end{equation}
with {\it critical value}
\begin{equation}
H_\alpha(y_c)=\frac{2a}{1+a}y_c=a y_c^{1+1/a},
\end{equation}
where $1/(1+a)^{1/\alpha}$ is a monotonically increasing real function from $0$ to 1 in $ (-1,\infty)$. (It is $1/e$ at $\alpha=0$) . Hence $y_c\le y_e$. Thus the critical value occurs in between $y=0$ and $y=y_e$.

\item[{\it iv)\ }]{\it The critical value is a maximum  in $(0,y_e)$ only if  $\alpha >0$.\ }

\item[{\it v)\ }] {\it The critical value is  minimum in  $(0,y_e)$  only if  $-1<\alpha <0$.\ } 

\item[{\it vi)\ }] {\it Stability of the period one orbits.\ }The fixed point $y=0$ is repelling for all $\alpha$;  $y_p=1$ is attracting for $ \alpha \in (0,2)$ and repelling for $a>2$.

\item[{\it vii)\ }] {\it Unbound orbits for $\alpha>0$.\ } As $\alpha$ increases past the threshold value $\alpha_t= 3.40349...$,the  critical value  $H_\alpha(y_c)$ becomes larger than the end point $y_e$ and the orbits escape to $-\infty$.

\end{itemize}

Based on these properties we conclude that the interesting interval of $\alpha$ is $(0,\alpha_t)$ since in this interval all orbits are bounded and  the period one fixed point $y_p=1$ loses its stability at $\alpha=2$. What remains to see is what kind of dynamics occurs from $\alpha=2$ to $\alpha_t$.

\begin{figure}[t]
\centerline{\includegraphics[width=0.95\columnwidth]{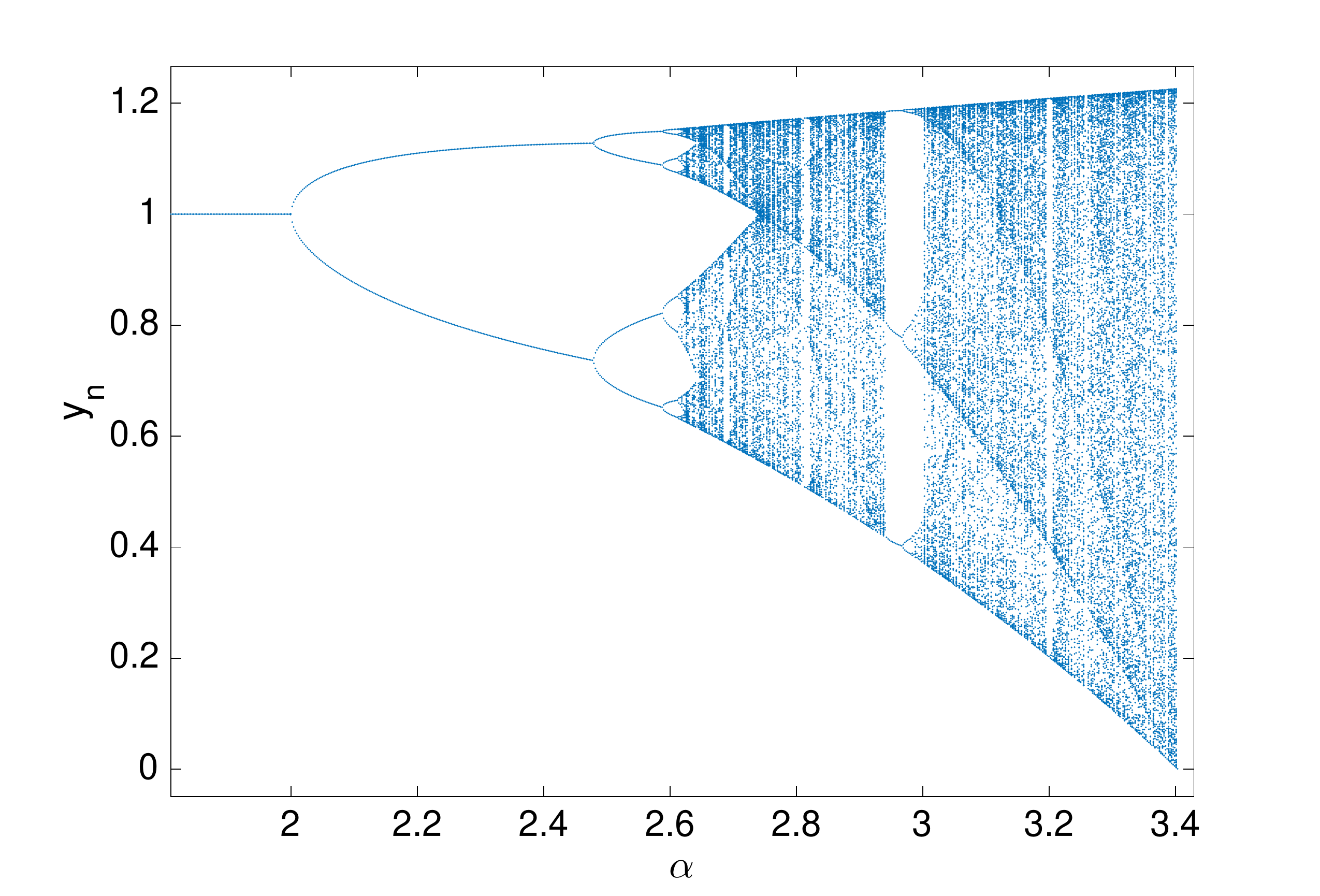}}
\caption{Bifurcation diagram for the simplified mapping (\ref{simplemap})}
\label{Bifur}
\end{figure}

\begin{figure}[t]
\centerline{\includegraphics[width=0.95\columnwidth]{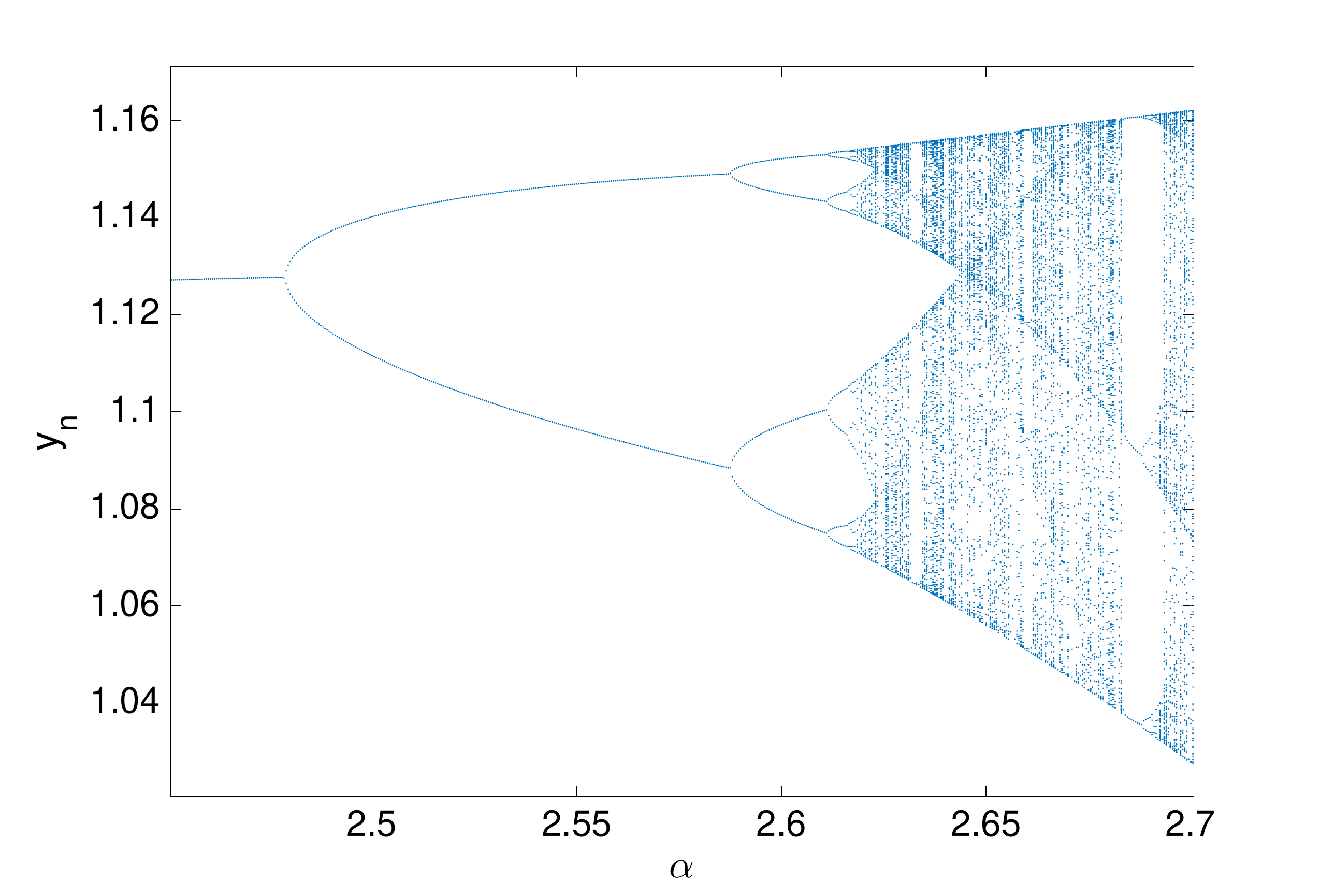}}
\caption{Zoom of Fig. \ref{Bifur}}
\label{BifurZoom}
\end{figure}

Figures \ref{Bifur} and \ref{BifurZoom} succintly display the rich dynamics  of the mapping $H_\alpha(y)$  in the interval $ 1.8\leq \alpha \leq \alpha_b$. The interval $ \alpha \in (0,1.8)$ is not included since all orbits  converge to the fixed point of period one $y_p=1$, as mentioned in item {\it vi} above. This bifurcation diagram shows not only the (predicted) instability of the period-one fixed point for $\alpha > 2$ but that the dynamics undergo the period doubling route to chaos. However, it should not be surprising given that  our simple mapping belongs to the universal class of unimodal mappings   with negative Schwarzian derivative in the interval $[0,y_e]$.\cite{Coullet} \cite{Guckenheimer}. 
   
\begin{figure}[t]
\centerline{\includegraphics[width=0.95\columnwidth]{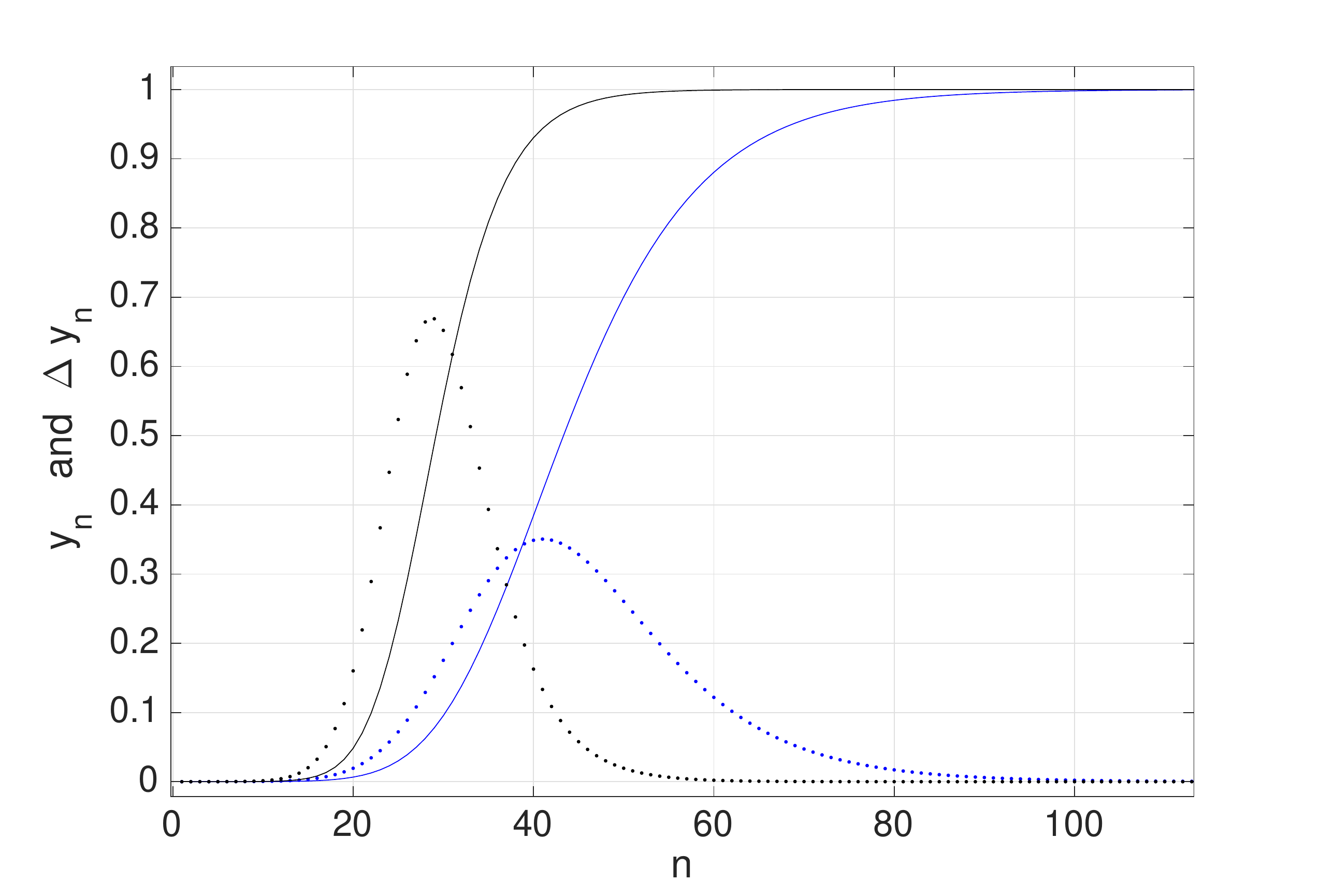}}
\caption{Temporal evolution of cummulative infections  $y_n $ vs.$n$ (solid lines) and daily new cases $\Delta y_n$ ( dotted lines), for two values of  $\alpha: \alpha=0.1$ ( blue curves) and $\alpha=0.2$ (black curves). Note that the plots for $\Delta y_n$ were multiplied by a factor of 10 for better visibility in a single graph}
\label{twosigmoids}
\end{figure}

{\it Remarks} As regards  the implementation of this simple map to the epidemics, one may also fit the data to the orbit (cascade) $y_n= H_\alpha(y_n)$. Of course, we would want a scenario where the orbits converge to one single outcome, which implies that $\alpha$ should be less than $2$. Small values of $\alpha$, say $\alpha=0.2$ yield sigmoid-like evolutions, Fig. \ref{twosigmoids} shows the evolution of $y(n)$ and daily new cases $\Delta y_n=y(n+1)-y(n)$  for  two representative small values of $\alpha=0.2$ (blue lines and dots and $\alpha=0.2$  (black lines and dot) with same initial conditions, $x(1)=10^{-6}$. \\
{\it The two parameter model $F(x,\alpha, r)$= Eq.(\ref{RichFlow}).}  A similar analysis can be carried out for the two parameter mapping (\ref{RichFlow}). It can be shown that the   maximum occus in three regions in parameter space, namely, region 1) $r>0, \alpha<-1$; region 2) $r<0, -1<\alpha<0$ and region 3) $r>0, \alpha>0$. Calculations of the Schwarzian derivative  of $F(x,\alpha, r)$  show that it is negative in large regions where the critical value of $F(x,\alpha, r) $ is a maximum, so it is expected that it will also display the period-doubling route to chaos. The detailed  analysis of this model shall be reported elsewhere.


\section{Discussion \label{sec:3}}

In this paper we have provided integrability conditions and exact solutions for a suffciently simple model of SIR. In small steps, we improved the model so as to encompass the variation of parameters with time and the interaction between regions, including migration of susceptible and infected individuals. A system with more than two regions is non-separable in any of its non-trivial versions; the exceptions, of course, are the absence of population exchange or the absence of infection. A variety of effects was discussed for infection revival under migratory conditions and variable quarantine policies.  

With respect to applicability, some comments are in order. In particular, Richards model (a generalized logistic model) was employed to fit Mexico's COVID-19  data up to a certain date of the epidemic. As a result, a variety of scenarios were obtained, consistent with strong fluctuations in the final number of casualties. It was possible to find a family of curves fitting the data up to a certain date reasonably well, but with strong variations in the final outcome, months ahead.
In this respect, we have seen how small variations in parameters produce large deviations in projected number of accumulated cases, and therefore, using $D = qR$, also in the number of casualties. This exhibits the challenge of reliable metrics for SIR and logistic type systems; a great deal of work and an impressive display of modelling methods in real time has been seen during the year 2020. To name a few important examples, we refer the reader to \cite{lancet1}, \cite{lancet2}, \cite{lancet3}. However, while acknowledging the merit  of these valuable works, we should underscore the fact that forecasting and nowcasting are strongly challenged by the careful determination of phenomenological parameters, especially the reproduction number $R_0$. Small inaccuracies may lead to very large errors, not because of chaoticity (which is absent) but because of exponential growth and the subsequent exponential depletion. 

In this respect, the existence of exact solutions allows for a systematic study of robustness, understood as error propagation under model variation and uncertainties. Interestingly, in the branch of quantum dynamics and complex modelling applied to physical systems, the concept of fidelity is sometimes mentioned as a useful measure, as it consists of an autocorrelation function between solutions after evolution takes place under two slightly different dynamical systems. We believe that a similar analysis can be carried out in population dynamics with the results provided in the present work.

Finally, it is important to stress the role of inverse problems, as opposed to direct problems, typically addressed in this setting. The former consists in the determination of a model from a desired behaviour, whereas the latter deals with the determination of the evolution from a given model. Inverse problems tap into exactly solvable systems and explicit formulas because of invertibility, a key concept guaranteed by the implicit function theorem.

%\section*{Declarations}
%%All manuscripts must contain the following sections under the heading 'Declarations'.
%%
%%If any of the sections are not relevant to your manuscript, please include the heading and write 'Not applicable' for that section.
%%
%%To be used for non-life science journals
%
%\begin{itemize}
%
%\item[] {\bf Funding\ } There was no funding specific for this work.\\
%
%
%\item[] {\bf Conflicts of interest/Competing interests\ } Not applicable.\\
%
%
%\item[] {\bf Availability of data and material\ } Not applicable. Public data quoted in References.\\
%
%
%\item[] {\bf Code availability\ } Not applicable.\\
%
%
%%\item[] {\bf Authors' contributions\ } Both authors conceived the project, wrote the manuscript and revised mathematical and numerical calculations.\\
%

%To be used for life science journals + articles with biological applications
%
%Funding (information that explains whether and by whom the research was supported)
%
%Conflicts of interest/Competing interests (include appropriate disclosures)
%
%Ethics approval (include appropriate approvals or waivers)
%
%Consent to participate (include appropriate statements)
%
%Consent for publication (include appropriate statements)
%
%Availability of data and material (data transparency)
%
%Code availability (software application or custom code)
%
%Authors' contributions (optional: please review the submission guidelines from the journal whether statements are mandatory)
%
%\end{itemize}

%\section*{References}

%\bibliography{mybibfile}

\end{document}